\documentclass[showpacs,prl,floatfix,twocolumn,amsmath,a4]{revtex4}
\usepackage{graphicx}

\begin{document}

\title{CdV$_2$O$_4$: A rare example of a collinear multiferroic spinel}

\author{G.~Giovannetti$^{1,2}$}
\author{A.~Stroppa$^{3}$}
\author{S.~Picozzi$^{1}$}
\author{D.~Baldomir$^{4}$}
\author{V.~Pardo$^{4}$}
\author{S.~Blanco-Canosa$^5$}
\author{F.~Rivadulla$^5$}
\author{S.~Jodlauk$^6$}
\author{D. Niermann$^7$}
\author{J. Rohrkamp$^7$}
\author{T.~Lorenz$^7$}
\author{J.~Hemberger$^{7}$}
\email[Corresponding author:~]{hemberger@ph2.uni-koeln.de}
\author{S.~Streltsov$^7$}
\author{D.I.~Khomskii$^7$}

\affiliation{$^1$CNR-SPIN L'Aquila, Italy}
\affiliation{$^2$ISC-CNR, Dipartimento di Fisica, Universit\`a ``La Sapienza'', Roma, Italy}
\affiliation{$^3$CNISM-Dipartimento di Fisica Universit\`a degli Studi di
L'Aquila, Italy }
\affiliation{$^4$Departamento de F\'{\i}sica Aplicada, Universidad de Santiago de Compostela, E-15782 Santiago de Compostela, Spain}
\affiliation{$^5$Departamento de Quimica-Fisica, Universidad de Santiago de Compostela, E-15782 Santiago de Compostela, Spain }
\affiliation{$^6$Institut f\"{u}r Kristallographie, Universit\"at zu K\"oln, D-50937 K\"oln, Germany }
\affiliation{$^7$II.\ Physikalisches Institut, Universit\"at zu K\"oln, D-50937 K\"oln, Germany }

\begin{abstract}
By studying the dielectric properties of the geometrically frustrated  spinel CdV$_2$O$_4$,
we observe ferroelectricity developing at the transition into the collinear antiferromagnetic ground state. In this multiferroic spinel, ferroelectricity is driven by local magnetostriction and not by the more common scenario of spiral magnetism. The experimental findings are corroborated by ab-initio calculations of the electric polarization and the underlying spin and orbital order. The results point towards a charge rearrangement due to dimerization, where electronic correlations and the proximity to the insulator-metal transition play an important role.
\end{abstract}

\date{\today}

\pacs{71.45.Gm, 71.10.Ca, 71.10.-w, 73.21.-b}

\maketitle

Spinels form a big class of materials - probably as big as perovskites. There are many magnetic materials among them, with rich magnetic properties. In contrast to perovskites, however, there are practically no ferroelectrics in this class. Why spinels are such ``bad actors" as to ferroelectricity, is actually not clear; the frustrated nature of their $B$-site sublattice can possibly play some role \cite{Seshadri}.  Only recently, the magnetically-driven ferroelectricity was found in some spinels with spiral magnetic structures, the best known example being CoCr$_2$O$_4$ \cite{Tokura}, and there are reports of ferroelectric-like properties in ferromagnetic semiconductors HgCr$_2$S$_4$ or CdCr$_2$S$_4$ \cite{Hemberger} and in charge-ordered magnetite \cite{Advmat,Brink,PicozziFe3O4}.

Here, we report the discovery of magnetically-driven ferroelectricity in a ternary spinel with a collinear magnetic structure CdV$_2$O$_4$ -- apparently the first such example. We found this multiferroic behavior experimentally, confirmed the presence of polarization and estimated its magnitude theoretically, using ab-initio calculations.

\begin{figure}
\centerline{\includegraphics[width=1.0\columnwidth,angle=0]{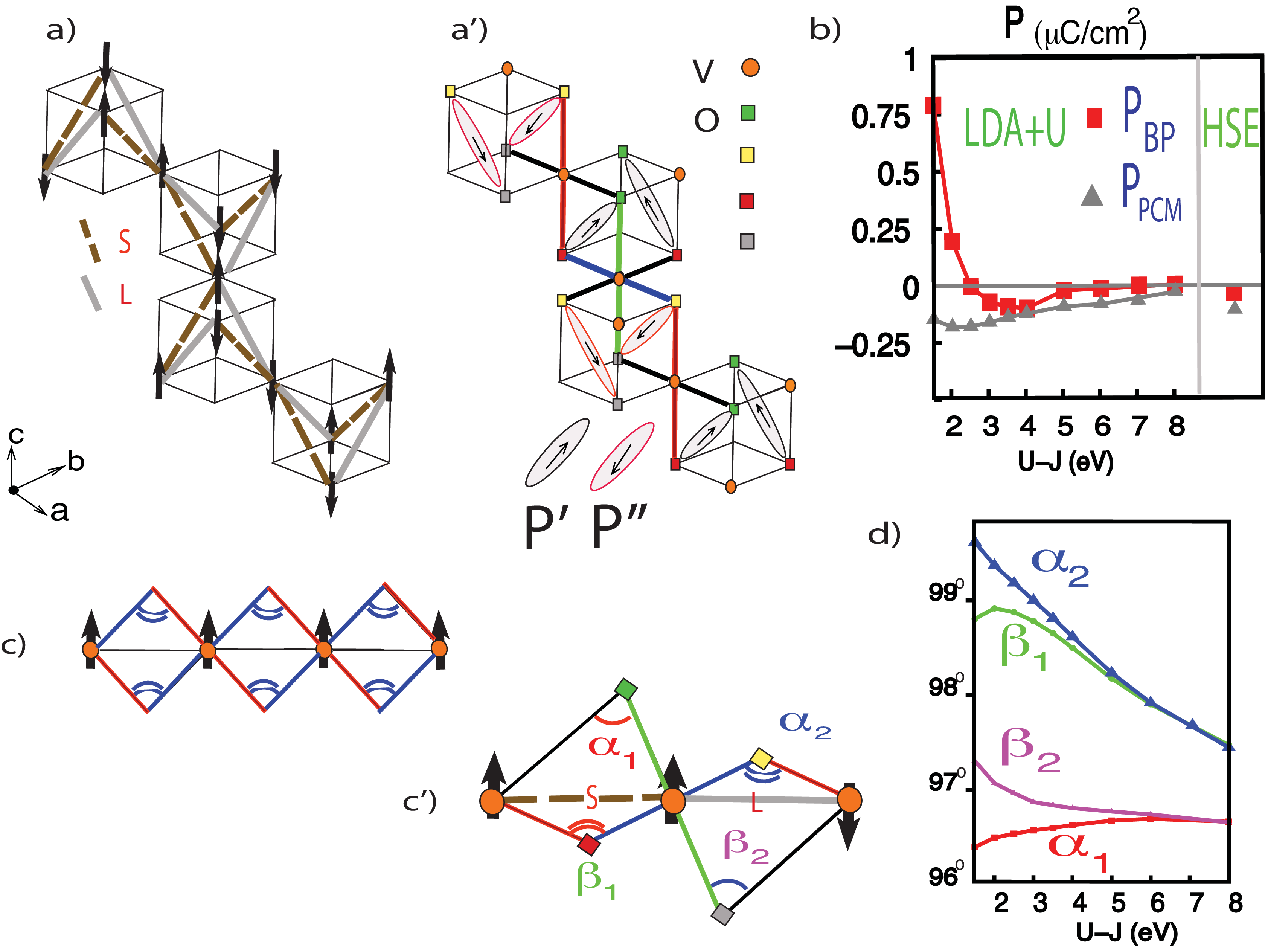}}
\caption{(Color online) Schematic view of a) the V-O ionic arrangements
with spin magnetic structure  (see black arrows) and short (long) V-V bond lengths highlighted by dashed (bold solid) lines;  a') oxygen-induced dipoles (see ellipses) and long bond-length V-O (see black lines) - symmetry inequivalent oxygens are denoted with different symbols/colors; b) calculated polarization according to  DFT+$U$ (left) and HSE schemes (right); V and O arrangement for c) $\uparrow\uparrow\uparrow\uparrow$ and c') $\uparrow\uparrow\downarrow\downarrow$ spin ordering
along [101] or [011]; d) values of ${\alpha}_1$, ${\beta}_2$, ${\alpha}_2$, ${\beta}_1$ angles as a function of $U_{eff}$.}
\label{theo}
\end{figure}

Vanadium spinels $A$V$_2$O$_4$ ($A$=Cd, Zn, Mg) recently attracted a lot of attention due to their  magnetic structure and due to a possibility of orbital ordering.
When  decreasing the temperature, all three materials experience  a structural transition from cubic to tetragonal  symmetry
\cite{Ueda}, and at lower temperatures they also show magnetic ordering. The magnetic structure is antiferromagnetic in $ab$ (or $xy$) chains, but is $\uparrow\uparrow\downarrow\downarrow$
in $xz$- and $yz$-chains (see Fig.~\ref{theo}).

Properties of related systems,  with  magnetic $A$-sites, \textit{e.g.}\ Mn or Fe, are similar, but the magnetic structures are more complicated. Several suggestions were proposed to explain the structural and magnetic transitions of V-spinels by different types of orbital ordering. V$^{3+}$ ions have two electrons in triply-degenerate $t_{2g}$ orbitals, which in the tetragonal phase are split into the lower occupied $xy$-orbital, and doubly-degenerate ($zx$, $yz$)-orbitals for the remaining second electron. At least three pictures of orbital ordering of this remaining electron were suggested: alternation of $xz$- and $xy$-orbitals in $z$-direction \cite{Motome}; occupation of complex orbitals $xz \pm i yz$ \cite{Tchernyshev}, and tetramerization (occupation $xz$-$xz$-$yz$-$yz$ in $xz$- and $yz$-chains) \cite{Mizokawa}. The description of orbital order should take into account the interplay between electron correlation, spin-orbit coupling, and cooperative JT distortions \cite{Valenti}.
Recently, a novel picture was suggested on the basis of ab-initio calculations \cite{Pardo}: almost no orbital ordering of this second electron, but strong alternation of V-V distances in $xz$- and $yz$-direction, with ferromagnetic bonds becoming  shorter.

The  magnetic structure, with $\uparrow\uparrow\downarrow\downarrow$ spin ordering along a chain, is reminiscent of the situation in E-type manganites like HoMnO$_3$ \cite{Dagotto,Lorenz} and in Ca$_3$CoMnO$_6$\cite{Cheong}, which are both multiferroic. This would suggest that multiferroicity can be found in V-spinels as well. This is indeed confirmed by our experimental and theoretical study.

\begin{figure}
\centerline{\includegraphics[width=0.85\columnwidth,angle=0]{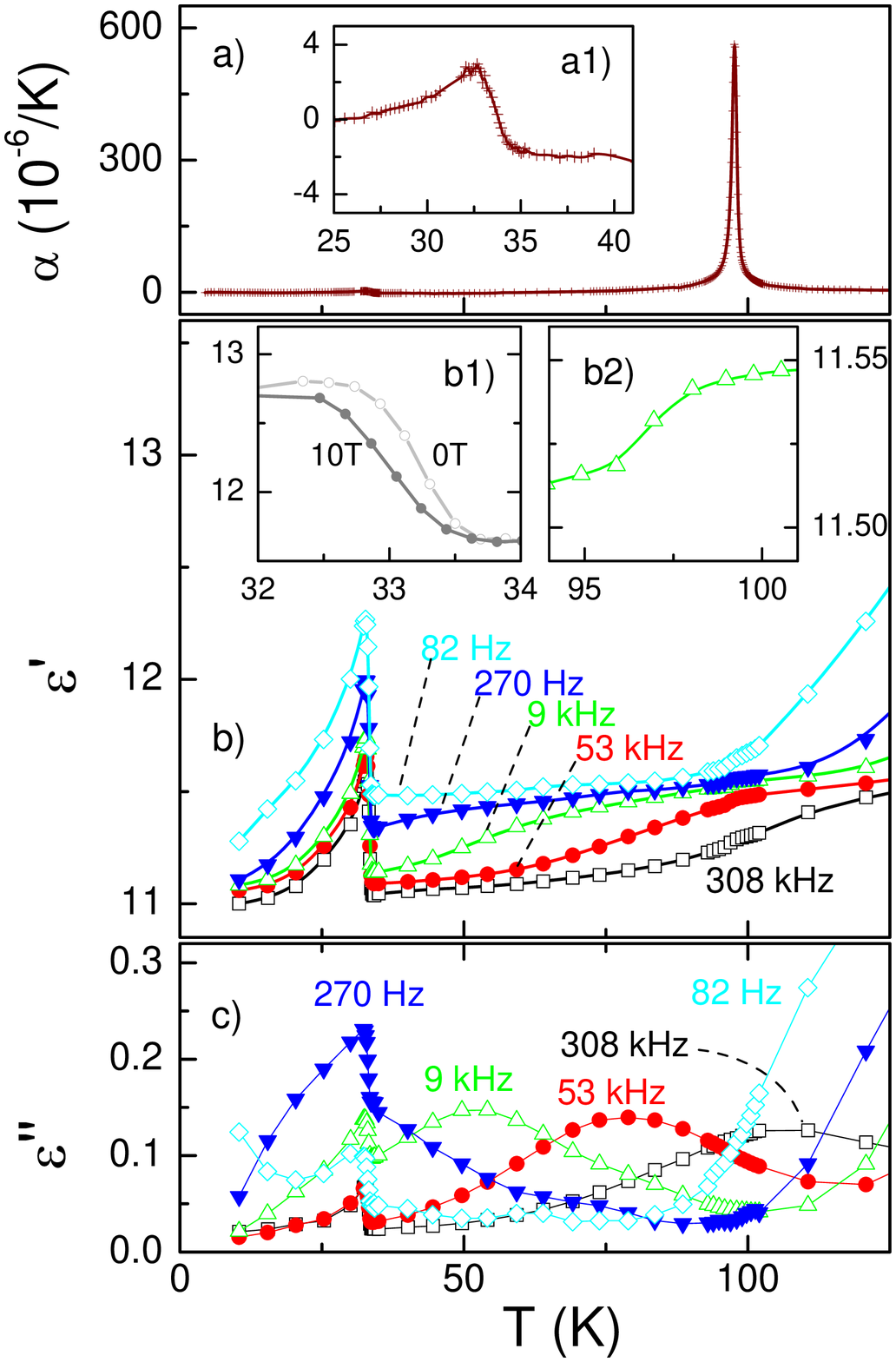}}
\caption{
(Color online) Thermal expansion $alpha$ in the whole temperature range (a) and in the vicinity of the magnetic transition (a1). The complex dielectric permittivity $\varepsilon^*$ of CdV$_2$O$_4$ is displayed as real part $\varepsilon'$ (b) and as dielectric loss $\varepsilon''$ (c) for different frequencies between 82~Hz and 0.3~MHz measured with a stimulus of $\sim$1~V/mm. The data displayed in frame b1) was measured in the vicinity of the orbital order transition at 1~Hz in external fields of zero and 10~T. Frame b2) shows the small dielectric anomaly at the structural transition below 100~K.
}
\label{figfreqinset}
\end{figure}

The single-phase, poly-crystalline samples of CdV$_2$O$_4$ were prepared by solid-state reaction in evacuated quartz ampoules as described in \cite{Goodenough}. Simultaneously, also the samples of ZnV$_2$O$_4$ and MgV$_2$O$_4$ were made. Structural and magnetic measurements confirmed the known behavior: in our samples of CdV$_2$O$_4$ a cubic-tetragonal transition occurs at $T_S \approx 95$~K, and at $T_N \approx 33$~K there appears antiferromagnetic ordering.
Both transitions cause pronounced anomalies in the thermal expansion coefficient $\alpha(T)$, measured in a home-built capacitance dilatometer (see Fig.~\ref{figfreqinset}a).
The measurements of electric polarization were made in a conventional $^4$He-flow magneto-cryostat (Oxford) by evaluating the integrated pyro-current recorded with a high-precision electrometer. Also the complex, frequency dependent dielectric response was measured employing a frequency-response analyzer (Novocontrol) with pseudo four-point probing for the linear measurements under small stimulus and with two-point probing for the non-linear measurements in electric driving-fields up to 220~V$_{rms}$.
For both purposes, silver paint contacts were applied to the platelike poly-crystalline pellets in sandwich geometry with a typical electrode area of $A\approx 10$~mm$^2$ and a thickness of $d\approx1$~mm. The uncertainty in the determination of the exact geometry together with additional (but constant) contributions of stray capacitances in two-point probing results in an uncertainty in the absolute values for electric polarization and permittivity of up to 20~\%.

\begin{figure}
\centerline{\includegraphics[width=0.8\columnwidth,angle=0]{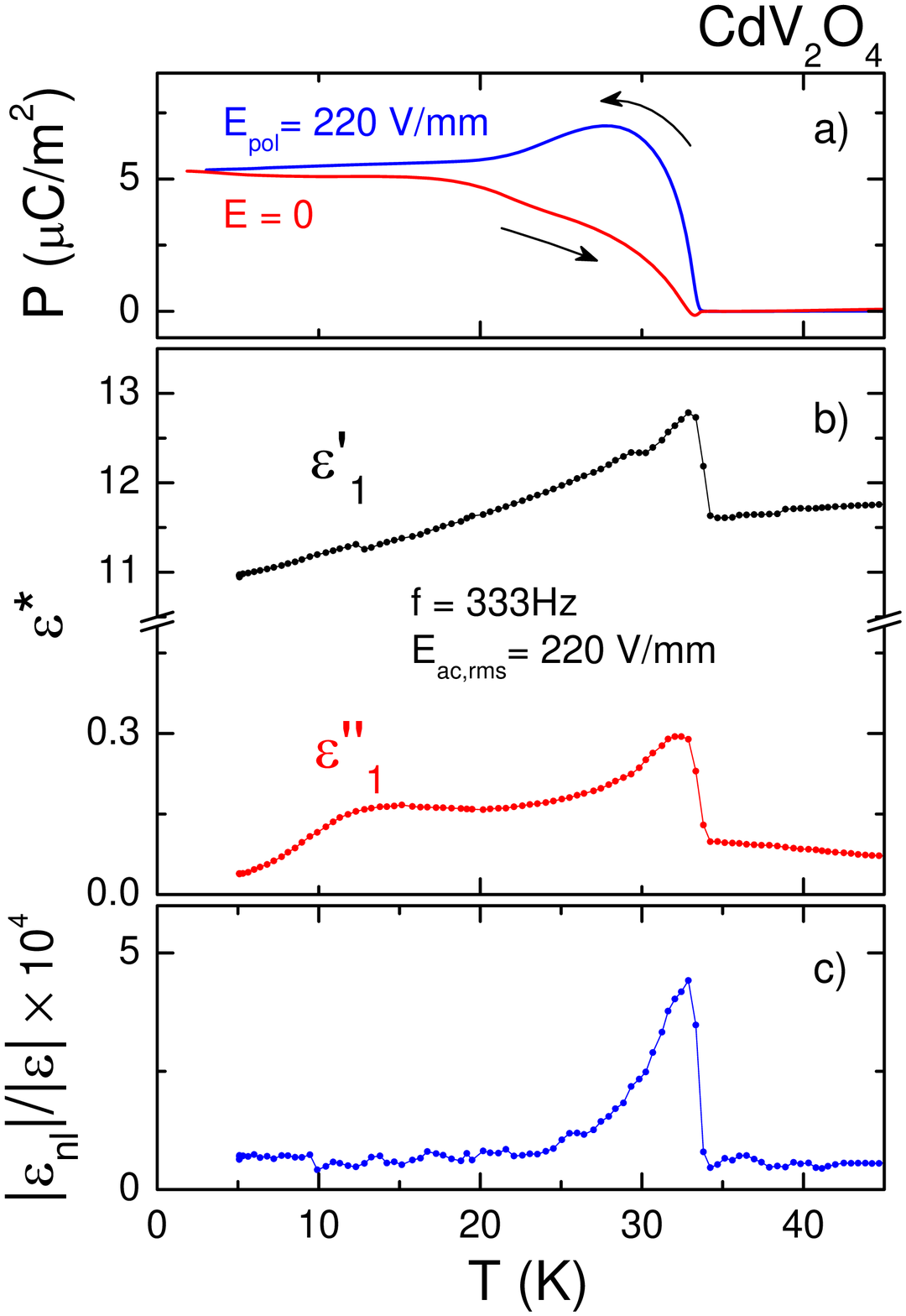}}
\caption{
(Color online) Electric polarization (a) measured via the pyro-current under cooling with a rate of 2~K/min in an electric poling field of 220~V/mm and under heating with the same rate in zero external field. Frame b) displays the linear component of the dielectric permittivity measured with a stimulus of 220~V$_{rms}$/mm. The corresponding (third order) nonlinear component $|\varepsilon_{nl}|$ normalized on the linear term is displayed in c); the "floor" of roughly $0.8\times10^{-4}$ can be attributed to the harmonic distortion of the high-voltage amplifier.
}
\label{p(E)}
\end{figure}

Results of the dielectric measurements are shown in Figs.~\ref{figfreqinset} and \ref{p(E)}.
The dielectric permittivity displays a small anomaly at the structural transition temperature $T_S\approx95$~K (Fig.~\ref{figfreqinset}b2) and a sharp jump-like anomaly at the magnetic transition $T_N\approx 33$~K. The latter slightly depends on an external magnetic field as demonstrated by the data for zero field and 10~T displayed in Fig.~\ref{figfreqinset}b1: the anomaly shifts slightly to lower temperatures under applying a magnetic field. At the same time, the permittivity possesses a distinct frequency dependent contribution superimposed on the described features: a step in the $\varepsilon'(T)$ is accompanied by a peak in $\varepsilon''(T)$ shifting to lower temperatures with decreasing frequency. A very similar relaxational contribution along the $c$-axis was found in multiferroic rare earth manganites. There, it was ascribed either to the freezing of an overdamped polar lattice-mode or, alternatively, one could think of the relaxation of localized polarons at defect states \cite{Schrettle}. In any case these data demonstrate that the dielectric properties are to some extent influenced by the experimental time window. 

That the sharp dielectric anomaly at $T_N$ is connected with the onset of spontaneous electric polarization is demonstrated in Fig.~\ref{p(E)}a displaying the electric polarization $P(T)$. The difference between the cooling data detected in the presence of an electric poling field, compared to the zero-field heating data, may be explained by field induced contributions like reversible domain orientation or the above mentioned relaxational contribution. The latter can be noticed as a broad loss peak in $\varepsilon''(T)$ above 10~K in the plot of the linear component of the complex permittivity, which were measured with a large stimulating electric field and is displayed in Fig.~\ref{p(E)}b. The behavior of the nonlinear component of the permittivity $\varepsilon_{nl}$ evaluated from the third harmonic of the dielectric response can be taken as an additional evidence for the onset of ferroelectricity (Fig.~\ref{p(E)}c). The onset of the third-order nonlinearity is followed by a decay towards lower temperatures which is related to the increase of the coercitivity: for temperatures below 25~K, the driving field of 220~V$_{rms}$/mm cannot switch ferroelectric domains anymore. It has to be noted that the second harmonic contribution (not shown) stays within the harmonic distortion of the setup at $T_N$, as expected for a symmetric polarization response $M(H)$.

All these data unequivocally demonstrate that there appears spontaneous polarization in CdV$_2$O$_4$ below $T_N$. The magnitude of polarization obtained in our samples (Fig.~\ref{p(E)}a) is $P \approx 5$~$\mu$C/m$^2$. One should expect that in a single-domain single crystal the polarization should be much larger. Similar measurements on poly-crystals of ZnV$_2$O$_4$ and MgV$_2$O$_4$ did not show ferroelectric behavior: polarization is absent (within the accuracy of $P < 0.1$~$\mu$C/m$^2$). We discuss possible reasons for such a different behavior below.

The explanation of the appearance of polarization in CdV$_2$O$_4$ has some analogy with the mechanism proposed for E-type manganites \cite{Dagotto,jcpslv}. The distortion, in particular the shifts of oxygens, are different in the vicinity of ferromagnetic bonds and antiferromagnetic ones. The main feature of the corresponding lattice distortions is the dimerization of V-V bonds \cite{Pardo}, but the shifts of the oxygens are also present, which ultimately lead to polarization. A specific feature of spinels, \textit{i.e.} the existence of {\em two} oxygens ``attached" to each V-V bond, compared to {\em one} oxygen in E-type manganites, makes the picture more complicated, but the main physical mechanism is very similar.


In order to shed light into the origin
of the ferroelectricity, we performed ab-initio calculations using the projector augmented-wave (PAW) method \cite{PAW}
with plane-wave basis sets as implemented in the VASP program \cite{VASP}. To describe  the
correlation effects of V-$3d$ electrons, we used the DFT+U method \cite{Dudarev}, 
for different values of $U_{eff}=U-J_H$ ranging between 0 and 8 eV, and the Heyd-Scuseria-Ernzerhof (HSE) 
hybrid functional \cite{hse}. The HSE  has been shown to provide an accurate treatment of  solid state systems when the  delicate balance between itinerancy and localization of correlated states plays an important role \cite{HSE1}, as in the system under study \cite{Pardo}.

Starting from the tetragonal experimental crystal structure with $I4_1/amd$ \textit{centric} symmetry \cite{CdV2O4expstr}, we performed  ionic relaxations for different values of $U_{eff}$, keeping the V magnetic moments ordered as $\uparrow\uparrow\downarrow\downarrow$ along the [101] and [011] directions (see Fig.~\ref{theo}a). The relaxed structure shows the formation of short (S) and long (L) bonds between $\uparrow\uparrow$ and $\uparrow\downarrow$ moments, respectively, see also \onlinecite{Pardo}. The situation seems to be different in MgV$_2$O$_4$ \cite{BelaLake}. Our relaxed structure has a $I4_1$ \textit{polar} space group.
For large values of $U_{eff}$, the general trend towards formation of  V-V dimers is reduced.
Note that, imposing the ferromagnetic spin ordering, the structure retains the inversion symmetry, and no dimerization is found for any values of $U_{eff}$. Further inspection of the relaxed polar ionic structures shows  that the spin, orbital and lattice degrees of freedom are coupled: in plane $xz,yz$ long bonds are oriented as reminiscent of a Kugel-Khomskii model \cite{K-K} and of the experimental structure proposed for ZnV$_2$O$_4$ \cite{Lee} (see Fig.~\ref{theo}a).
For the relaxed structure, we evaluated the ferroelectric polarization $P_{BP}$  (see Fig.~\ref{theo}b) according to the Berry phase theory \cite{BerryPhase}. We obtained a finite $P_{BP}$, directed along the $c$-axis, which is strongly related to the formation of a dimerized structure at intermediate values of $U_{eff}$.
The electronic charge redistribution due to dimer formation gives rise to differences between  the BP and the point charge model (PCM) values  of $P$.
The red curve in Fig.~\ref{theo}b) is the total polarization, ionic plus electronic (BP) contribution, \textit{i.e.} $P_{BP}$, the grey one is the estimate of the point charge model (using the ionic nominal valencies). The fact that they are different suggests that covalency effects are relevant.

We now discuss the mechanism behind ferroelectricity. First of all, let us recall that if we consider a FM ($\uparrow\uparrow\uparrow\uparrow$) state  with the suggested experimental $I4_1/amd$ symmetry\cite{CdV2O4expstr},
all V-O-V angles are equivalent along the chain (see Fig.~\ref{theo}c).
Note, however, that V-O distances are slightly different, due to the peculiar coordination of the spinel structure: each O is an ``apical" one with respect to one V ion and a ``planar" one with respect to the other neighboring V ion. This does not exclude a priori some degree of orbital-ordering, even in a FM spin configuration. As expected from the centrosymmetric space group, no polarization is found from our calculations for this case.
However, once the magnetic order $\uparrow\uparrow\downarrow\downarrow$
is imposed, (i) the angles ${\alpha}_1$ and ${\beta}_2$ (${\alpha}_2$ and ${\beta}_1$) become inequivalent as a consequence of the formation of short/long V-V bonds; (ii)  ${\alpha}_1$ and ${\beta}_1$ become different. The long V-O bonds are arranged as shown in Fig.~\ref{theo}a'. This pattern is compatible with a weakly staggered \textit{xz},\textit{yz} orbital ordering. As a result, the dipole moments P$^{'}$ and P$^{''}$, originating from the inequivalency of oxygens, and  schematically shown in Fig.~\ref{theo}a), appear due to different ${\alpha}_1$
and ${\beta}_2$ (${\alpha}_2$ and ${\beta}_1$) angles; since P$^{'}$ and P$^{''}$ don't compensate, they give rise to a net $P$. This picture is valid for intermediate values of $U_{eff}$. At large values of U$_{eff}$, however, the inequivalence of ${\alpha}_1$ and ${\beta}_2$ (${\alpha}_2$ and ${\beta}_1$) V-O-V angles disappears , see point (i) above, whereas (ii) is still valid. 
(see Fig.~\ref{theo}d); as a result, $P$ is suppressed, but is still non-zero.
In other words,  the $\uparrow\uparrow\downarrow\downarrow$ spin ordering, imposed onto the centric $I4_1/amd$ space group, gives rise to an \textit{electronic instability} which ultimately results in a V-V dimerization \textit{and} formation of short and long V-O bonds, compatible with a staggered \textit{xz},\textit{yz} orbital ordering. This electronic instability is already evident \textit{before} ionic optimization:  the  two $\uparrow$ ($\downarrow$) V sites are inequivalent by symmetry, and, upon ionic  relaxations, this inequivalency eventually drives the V-V dimerization. Also  the two oxygens  bonded to $\uparrow$,$\uparrow$ V (or $\downarrow$,$\downarrow$ V) become inequivalent, in turn giving rise, upon ionic relaxations,
to a weakly staggered orbital ordering.
Both effects cooperate to stabilize the polarization.


Thus the theoretical results  confirm, first of all, the structure proposed in \cite{Pardo}, with dimerization in V-V chains in $xz-$ and $yz$-directions. Furthermore, they show that   CdV$_2$O$_4$ is  ferroelectric. The calculated polarization is along the $c$-direction, and its value is  $P\approx 200$~$\mu$C/m$^2$. This is typical for multiferroics with ferroelectricity  induced by magnetostriction, like HoMnO$_3$  or YMn$_2$O$_5$, see e.g.\ \cite{Lorenz}. The fact that the experimentally observed value is smaller, $P\approx 5$~$\mu$C/m$^2$, see above, is rather common. The same happens in HoMnO$_3$: the measured value in polycrystalline material $P\approx90$~$\mu$C/m$^2$ \cite{Lorenz} is smaller than the theoretical value \cite{Dagotto} - which however agrees with the estimates obtained from optical studies \cite{Mostovoy-Aguilar}. Most likely it is connected with the poly-crystalline nature of the samples: first, polarization is averaged over all directions, but most important is that probably in these granular materials one does not reach full domain orientation during the poling procedure used.

Why the polarization is absent in the measured samples of ZnV$_2$O$_4$ and MgV$_2$O$_4$, is not completely clear. Theoretically we could expect that these spinels, with very similar structure, could be also multiferroic. One of the reasons could be that these materials have much smaller gaps than CdV$_2$O$_4$ and are close to the localized-itinerant crossover \cite{Goodenough}. The finite conductivity of these samples (10$^4$ times higher than in CdV$_2$O$_4$ \cite{Pardo}) may scramble the results of the measurements. The quality of the samples may also matter. In any case, this question deserves further study.

Summarizing, we discovered the first multiferroic ternary spinel with collinear magnetic structure, CdV$_2$O$_4$. We thus show that not only a spiral magnetic structure can produce ferroelectricity in spinels; the magnetostriction mechanism can do it as well. In the latter case, polarization is usually even larger than in spiral magnets, which is also the case here.
Our study also clarifies the very controversial question about orbital ordering and structural distortions in V-spinels, confirming that there should appear a strong V-V dimerization, that is also responsible for polarization.  We conclude by suggesting that similar phenomena might occur in other spinels, thus broadening the class of multiferroic systems to this important group of materials.


The research leading to the ab-initio results has received funding from the European Research Council under the EU 7th Framework Programme (FP7/2007-2013)/ERC Grant Agreement No.203523. The experimental work has been funded by the DFG through SFB608 (Cologne).


\end{document}